%% file: IZS_seidl.tex
\documentclass[a4paper,conference]{IEEEtran}

\usepackage{tikz}
\usepackage{pstricks}
\usepackage{psfrag}

\usepackage{url}
\usepackage{ifthen}
\usepackage[cmex10]{amsmath} 
\input{makros}

\begin{document}
\title{An Efficient Length- and Rate-Preserving Concatenation of Polar and Repetition Codes}

\author{%
\IEEEauthorblockN{Mathis Seidl and Johannes B. Huber}
\IEEEauthorblockA{Lehrstuhl f{\"u}r Informations{\"u}bertragung\\
              Universit{\"a}t Erlangen-N{\"u}rnberg\\
              Erlangen, Germany\\
              Email: {\tt \{seidl, huber\}@LNT.de}}
}
\maketitle

\begin{abstract}
  We improve the method in~\cite{Seidl:10} for increasing the finite-lengh performance of polar codes by 
  protecting specific, less reliable symbols with simple outer repetition codes. 
  Decoding of the scheme integrates easily in the known successive decoding algorithms for polar codes. 
  Overall rate and block length remain unchanged, the decoding complexity is at most doubled. 
  A comparison to related methods for performance improvement of polar codes is drawn. 
\end{abstract}

\section{Introduction}
\label{sec:intro}
\input{sec_intro}

\section{Polar Codes and their Decoding}
\label{sec:polar}
\input{sec_polar}

\section{Concatenated Code Construction}
\label{sec:inner}
\input{sec_inner}
\section{Simulation Results}
\label{sec:sim}
\input{sec_sim}
\section{Conclusion}
\label{sec:conc}
\input{sec_conc}

\vspace*{6.5mm}
\bibliographystyle{IEEEtran}
\bibliography{mjssBib}


\end{document}

%% file: makros.tex
\def\ve#1{{\mathchoice{\mbox{\boldmath$\displaystyle #1$}}%
              {\mbox{\boldmath$\textstyle #1$}}%
              {\mbox{\boldmath$\scriptstyle #1$}}%
              {\mbox{\boldmath$\scriptscriptstyle #1$}}}}
\DeclareSymbolFont{AMSb}{U}{msb}{m}{n}
\DeclareSymbolFontAlphabet{\mathbb}{AMSb}

\def\e{\mathrm{ e}}

\def\Pr{\mathrm{Pr}}

\newcommand {\Es} {E_{\mathrm{s}}}            
\newcommand {\Eb} {E_{\mathrm{b}}}     
\def\dB{\mathrm{dB}}

\def\No{N_0}

\newcommand{\argmax}{\mathop{\mathrm{argmax}}}

%% file: sec_intro.tex
Polar coding is known as a channel coding construction that is able to achieve the capacity of many symmetric discrete memoryless channels under low-complexity $\mathcal O(N\log N)$ encoding and successive decoding~\cite{Arikan:09}. Unfortunately, the error performance of polar codes for finite block lengths is quite moderate. The key feature of polar coding -- when compared to other existing channel block coding schemes -- clearly lies in its low decoding complexity. Therefore, the trade-off between computational complexity and error performance for polar codes is of interest, i.e., the development of efficient methods that allow for better performance at moderate additional complexity. 

Optimizing the decoding algorithm for polar codes has been the subject of various work, e.g.~\cite{TalVardy:11, Kschischang:11} and has led to substantial improvements. 
Though, apart from the decoder, the code itself leaves room for improvement as well. 

To this end, we propose a modified polar code construction by means of a serial concatenated scheme with the polar code 
used as an inner code. 
In contrast to many existing concatenation schemes based on polar codes as inner codes (as considered, e.g., in~\cite{Bakshi:10, Seidl:11, Mahdavifar:13}), 
we focus here on coding schemes that do not change overall rate and block length, thus facilitating a pure 
trade-off of complexity and error performance. 
The approach is based on our prior attempt~\cite{Seidl:10} where block codes of small dimension were chosen as outer codes. 
In this paper we show that -- by an efficient and systematic design -- 
an equivalent, significant performance gain is achieved by protecting an inner polar code 
with only one-dimensional outer codes, i.e., repetition codes,  resulting in an quite smaller increase of complexity. 
Furthermore, we relate the results to methods where the decoder is modified instead of the code. 

The paper is organized as follows: After a brief review on polar codes and their decoding strategies in Sec.~\ref{sec:polar}, 
we describe our concatenated code construction in Sec.~\ref{sec:inner}, followed by simulation results in Sec.~\ref{sec:sim} and 
some conclusive remarks in Sec.~\ref{sec:conc}.

%% file: sec_polar.tex
\subsection{Code Construction}
  Since the concept of polar coding is widely known, we only give a brief overview, focussing 
  on the aspects of importance for this paper. 
  We follow the original approach in~\cite{Arikan:09} where the generator matrix 
  is chosen as a subset (indexed by $\mathcal A$) of the rows of the binary matrix
  \begin{equation} 
    \ve{G}_N = \ve{F}^{\otimes n} \quad , \quad \ve{F} = \begin{bmatrix}1&0\\1&1\end{bmatrix}
  \end{equation} 
  with $n=\log_2 N$ and $\otimes n$ denoting the $n$-th Kronecker power. 
  
  Under successive decoding, the transmission of the particular source symbols $u_i$ may be described by their own binary-input 
  channels (\emph{bit channels}) which show a polarization effect in the sense that their capacities are almost all either near $0$ or near $1$. 
  These capacities -- or equivalently, the corresponding failure probabilities $p_\e(i)$) -- can be easily determined. 
  The channels with high capacities are chosen to form the set $\mathcal A$ whereas the residual channels (\emph{frozen channels}) transmit fixed 
  values that are known to the decoder. 
  
\subsection{Successive Decoding}
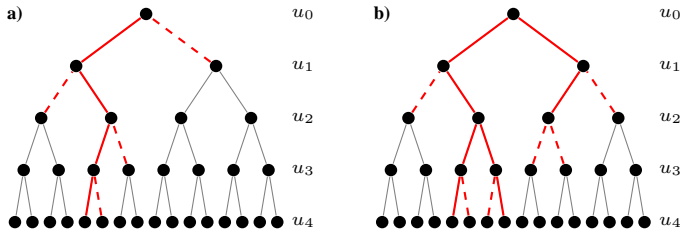
\begin{figure}[width=0.45\textwidth]
    \begin{centering} %
  \beginpgfgraphicnamed{pic_sc-external}%
  \input{pic_sc.tikz}%
  \endpgfgraphicnamed%
 \end{centering}
    \caption{\label{algo_sc_scl}a) SC decoding.\quad  b) Successive list decoding (list size $L=2$). Bold face lines: Inspected (and tentatively selected) paths in the SC decoding process. Dashed lines: 
              Inspected but discarded paths. Thin lines: Never inspected paths.}
  \end{figure}

  In the successive cancellation (SC) decoding approach~\cite{Arikan:09}, estimates $\hat u_i$ on the source symbols 
  $u_i$ ($i \in \mathcal A$) are calculated successively, according to the recursion formula
  \begin{equation}\label{scdec}
    \hat u_i := \argmax_{b \in \{0,1\}} \Big\{ \Pr \big(U_i = b | \ve{Y}, \hat U_0 \cdots \hat U_{i-1}\big) \Big\} \; .
  \end{equation}
  Thus, in each step $i$ the decoder checks which of the possible two values for $u_i$ is more likely, given the received vector $\ve{y}$ as well as the sequence $\hat u_0\cdots \hat u_{i-1}$ 
  of data symbols already decided in the previous steps. 
  Due to the special structure of $\ve{G}_N$, the calculation of the probabilities in \eqref{scdec} can be implemented in an FFT-like fashion, resulting in a low $\mathcal O(N\log N)$ 
  overall decoding complexity. 
  With increasing SNR, the performance of the SC algorithm is known to converge to that of an optimum Maximum-Likelihood (ML) decoder. 
  The decoding process as a path search is illustrated in Fig.~\ref{algo_sc_scl}a). 

  The word error performance under SC decoding can be precisely determined. It is given by the term
  \begin{equation}\label{sc_wer}
    \mathrm{WER}_{\mathrm{SC}} = 1 - \prod_{i \in \mathcal A} (1-p_\e (i) )
  \end{equation}
  where $p_\mathrm{e}(i)$ denotes the probability of a wrong decision at stage $i$ of the decoder provided that all previous decisions have been correct. 
  From~\eqref{sc_wer} it is clear that for an optimal code construction, $\mathcal A$ should consist of the bit channels with lowest 
  failure probabilities $p_\e(i)$. 

\subsection{Successive List Decoding}
  As an improved version of the SC decoder for increased performance in the low-SNR regime, list decoding for polar codes has been proposed~\cite{TalVardy:11}. 
  The successive list decoder does not take hard decisions on the $u_i$ immediately. 
  Instead, both possible values are examined in separate decoding branches, and the corresponding likelihood values are determined. 
  If the number of branches exceeds a certain design parameter $L$ (the \emph{list size}), the least probable branches are discarded, as examplary visualized in Fig.~\ref{algo_sc_scl}b) for $L=2$.
  The complexity of decoding scales linearly with the list size $L$ and is of order $\mathcal O(LN\log N)$. 
  Note that  only the decoder is modified here while the code does not change.

  In an extended version of the above-mentioned paper~\cite{TalVardy_arxiv:11}, the authors propose a serial concatenation scheme with an inner polar code and a very high-rate outer 
  CRC (cyclic redundancy check) code. 
  Decoding for this scheme is accomplished in two steps: First, the successive list decoder generates a list of $L$ possible codewords. After that, the CRC sums for each 
  entry of the list are calculated in order to check for the correct codeword. By this means, correct decoding is possible in principle even when another polar codeword in the list 
  belongs to a more likely path, enabling successful decoding beyond the 
  performance of an ML decoder for the inner polar code alone -- as long as the correct codeword is part of the output list.
  It has been shown~\cite{TalVardy_arxiv:11} that by this means, a significant performance gain is achieved. 

%% file: pic_sc.tikz
\begin{tikzpicture}
[lineDecorate/.style={-, color=gray},%
lineDecorate2/.style={-,thick,color=red},%
lineDecorate4/.style={-,dashed, thick,color=red},%
VarNode/.style={shape=circle,fill=black, thin, inner sep=0pt},%
CheckNode/.style={shape=rectangle,inner sep=2.5pt,draw,thick}, scale=0.23]
\scriptsize
\foreach \nodename/\nodelabel/\x/\y/\direction/\navigate in {
u40/c/0/0/bottom/south, u41/c/1/0/bottom/south, u42/c/2/0/bottom/south, u43/c/3/0/bottom/south, 
u44/c/4/0/bottom/south, u45/c/5/0/bottom/south, u46/c/6/0/bottom/south, u47/c/7/0/bottom/south, 
u48/c/8/0/bottom/south, u49/c/9/0/bottom/south, u410/c/10/0/bottom/south, u411/c/11/0/bottom/south, 
u412/c/12/0/bottom/south, u413/c/13/0/bottom/south, u414/c/14/0/bottom/south, u415/c/15/0/bottom/south, 
u30/c/0.5/3/bottom/south, u31/c/2.5/3/bottom/south, u32/c/4.5/3/bottom/south, u33/c/6.5/3/bottom/south, 
u34/c/8.5/3/bottom/south, u35/c/10.5/3/bottom/south, u36/c/12.5/3/bottom/south, u37/c/14.5/3/bottom/south, 
u20/c/1.5/6/bottom/south, u21/c/5.5/6/bottom/south, u22/c/9.5/6/bottom/south, u23/c/13.5/6/bottom/south, 
u10/c/3.5/9/bottom/south, u11/c/11.5/9/bottom/south, 
u00/c/7.5/12/bottom/south,
v40/c/21/0/bottom/south, v41/c/22/0/bottom/south, v42/c/23/0/bottom/south, v43/c/24/0/bottom/south, 
v44/c/25/0/bottom/south, v45/c/26/0/bottom/south, v46/c/27/0/bottom/south, v47/c/28/0/bottom/south, 
v48/c/29/0/bottom/south, v49/c/30/0/bottom/south, v410/c/31/0/bottom/south, v411/c/32/0/bottom/south, 
v412/c/33/0/bottom/south, v413/c/34/0/bottom/south, v414/c/35/0/bottom/south, v415/c/36/0/bottom/south, 
v30/c/21.5/3/bottom/south, v31/c/23.5/3/bottom/south, v32/c/25.5/3/bottom/south, v33/c/27.5/3/bottom/south, 
v34/c/29.5/3/bottom/south, v35/c/31.5/3/bottom/south, v36/c/33.5/3/bottom/south, v37/c/35.5/3/bottom/south, 
v20/c/22.5/6/bottom/south, v21/c/26.5/6/bottom/south, v22/c/30.5/6/bottom/south, v23/c/34.5/6/bottom/south, 
v10/c/24.5/9/bottom/south, v11/c/32.5/9/bottom/south, 
v00/c/28.5/12/bottom/south}
{ \node (\nodename) at (\x,\y) [VarNode] {$\nodelabel$}; }
\path
\foreach \startnode/\endnode in {
u40/u30, u41/u30, u42/u31, u43/u31, u44/u32, u46/u33, u47/u33, u48/u34, u49/u34, 
u410/u35, u411/u35, u412/u36, u413/u36, u414/u37, u415/u37, 
u30/u20, u31/u20, u32/u21, u34/u22, u35/u22, u36/u23, u37/u23, 
u21/u10, u22/u11, u23/u11, u10/u00}
{ (\startnode) edge[lineDecorate] (\endnode) };


\path
\foreach \startnode/\endnode in {u44/u32, u32/u21, u21/u10, u10/u00}
{ (\startnode) edge[lineDecorate2] (\endnode) };
\path
\foreach \startnode/\endnode in {u45/u32, u33/u21, u20/u10, u11/u00}
{ (\startnode) edge[lineDecorate4] (\endnode) };
\node (u0) at (16.5,12) {$u_0$};
\node (u1) at (16.5, 9) {$u_1$};
\node (u2) at (16.5, 6) {$u_2$};
\node (u3) at (16.5, 3) {$u_3$};
\node (u4) at (16.5, 0) {$u_4$};
\node (v0) at (37.5,12) {$u_0$};
\node (v1) at (37.5, 9) {$u_1$};
\node (v2) at (37.5, 6) {$u_2$};
\node (v3) at (37.5, 3) {$u_3$};
\node (v4) at (37.5, 0) {$u_4$};
\node (a) at (0, 12) {\bfseries{a)}};
\node (b) at (21, 12) {\bfseries{b)}};

\path
\foreach \startnode/\endnode in {
v40/v30, v41/v30, v42/v31, v43/v31, v44/v32, v47/v33, v48/v34, v49/v34, 
v410/v35, v411/v35, v412/v36, v413/v36, v414/v37, v415/v37, 
v30/v20, v31/v20, v32/v21, v33/v21,v36/v23, v37/v23, 
v21/v10, v22/v11, v10/v00, v11/v00}
{ (\startnode) edge[lineDecorate] (\endnode) };


\path
\foreach \startnode/\endnode in {
v44/v32, v32/v21, v21/v10, v10/v00, 
v11/v00, v22/v11, v33/v21, v47/v33}
{ (\startnode) edge[lineDecorate2] (\endnode) };
\path
\foreach \startnode/\endnode in {
v45/v32, v35/v22, v20/v10,  
v23/v11, v34/v22, v46/v33}
{ (\startnode) edge[lineDecorate4] (\endnode) };
\end{tikzpicture}

%% file: sec_inner.tex
Here, we follow a different approach based on the 
varying bit channel capacities under successive decoding. 
The proposed coding scheme follows the conventional serial concatenation principle where the source symbols are first encoded by an outer code, 
followed by an inner encoding. Thus, the overall rate is given as $R=R_\mathrm{o}R_\mathrm{i}$ with $R_\mathrm{o}$ and 
$R_\mathrm{i}$ being the rate of outer and inner code, respectively. 
In our approach, outer and inner code are decoded jointly by a single algorithm. 

The generator matrix $\ve{G}$ of a $(N,K)$ polar code constructed in the conventional way may be represented as 
\begin{equation}
  \ve{G} = \ve{P}_{\mathcal A} \cdot \ve{G}_N
\end{equation}
where $\ve{P}_{\mathcal A}$ is a $(K \times N)$ projection matrix with rows built from the $i$-th unit vectors
of length $N$ ($i\in \mathcal A$, $|\mathcal A|=K$).
We now aim to construct an optimized generator matrix of equal dimensions by a serial concatenation
of the form
\begin{equation}\label{conc_enc}
  \ve{G}^\ast = \ve{G}_{\mathrm{o}} \cdot \left( \ve{P}_{\mathcal A^\ast} \cdot \ve{G}_N \right)
\end{equation}
based on an enlarged set of channel indices $\mathcal A^\ast$ with $K<|\mathcal A^\ast|\leq N$. 
The $(K \times |\mathcal A^\ast |)$ matrix $\ve{G}_\mathrm{o}$ serves as a generator matrix of a suitably chosen outer code
\footnote{While the code is designed to operate \emph{inside} the successive decoding process, 
from the encoding procedure \eqref{conc_enc} it becomes clear that it serves in fact as an outer code. In our prior paper~\cite{Seidl:10}, the denotation 
''inner code`` had been used from the decoding perspective which indeed is misleading.}. 

In the following, we demonstrate the code construction by means of an example considering a rate-$1/2$, length-$256$ polar code. 

\subsection{Inner Code Design}
  \begin{figure}
    \psfrag{chindex}[Bc][Bc][0.8]{bit channel index $i\ \rightarrow$}
    \psfrag{perror}[Bc][Bc][0.8]{$p_\e (i)$}
    \centerline{\includegraphics[width=0.48\textwidth]{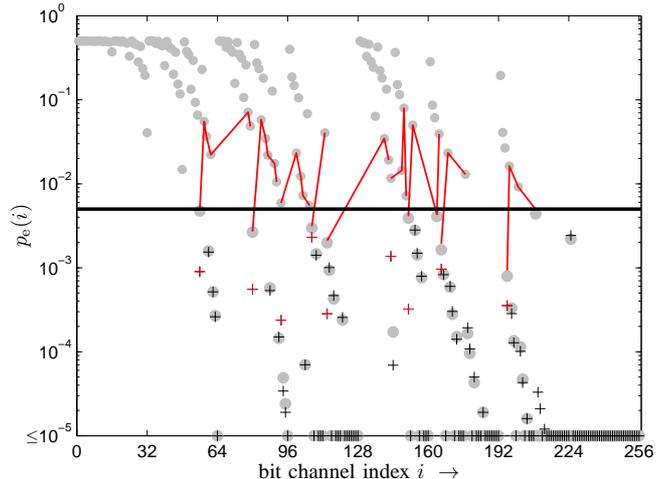}\rput(-8.5,0.65){\tiny$\leq$}}
    \caption{\label{fig_bitch}Failure probabilities $p_\e (i)$ for a polar code ($R=1/2$, $N=256$, BI-AWGN channel at $\Es / \No = -0.5\;\dB$). 
               Gray circles: original code. Markers: concatenated code. Red lines: Repetition blocks of outer code.}
  \end{figure}
  The gray circles in Fig.~\ref{fig_bitch} show the failure probabilities $p_\e(i)$  of the bit channels after transmission  over a binary-input AWGN channel at 
  $\Es / \No=-0.5\;\dB$. The black line corresponds to the design rate $R=1/2$ of the original code. The indices $i$ with $p_\e(i)$ 
  below this threshold form the set $\mathcal A$.  

  For construction of a concatenated code with equal rate and block length from a given $(N,K)$ polar code, 
  we choose the inner code as a polar code with same length $N$ but with a higher rate $R_\mathrm{i}>R$. 
  This is easily accomplished by enlarging the set $\mathcal A$ 
  of information symbols, i.e., using additional (previously frozen) bit channels for transmission 
  to form the set $\mathcal A^\ast$.

\subsection{Outer Code Design}
  As can be derived from \eqref{sc_wer} and Fig.~\ref{fig_bitch}, the word error rate is dominated by a comparatively small fraction of bit channels 
  close to the threshold. We now aim to protect these least-reliable bit channels by a suited outer code including some additional (formerly frozen) channels 
  that are provided by the inner enlarged polar code. 
   
  In contrast to our previous approach~\cite{Seidl:10}, in this paper we aim to  
  minimize the additional complexity introduced by outer decoding which imposes a number of constraints 
  on the code construction that are explained in the following:  
  
  First, we focus on simple one-dimensional codes, i.e., repetition codes. 
  Thus, the outer coding actually consists in setting some of the source symbols to the same value and 
  building small sets of combined channel indices from $\mathcal A^\ast$ (\emph{repetition blocks}). 
  For further complexity reduction, we require that these blocks do not overlap 
  (in the sense that the contained indices do not overlap). 

  The use of more than one bit channel for transmitting a single bit of information may be represented by a single equivalent bit channel (red markers 
  in Fig.~\ref{fig_bitch}). 
  We found that protecting (merging) two channels in this way always leads to an improvement w.r.t. the first-positioned channel, 
  but not necessarily when compared to the second one. This is due to the successive decoding strategy: 
  The decision on a repetition block is made at reaching the end of the block, like explained in detail in the next subsection. 
  If already at the first index of a block a wrong codeword corresponds to the more likely path, decoding of the following symbols (that are in general not protected 
  by the outer code) is quite likely to fail, even when both possible values for the first symbol are pursued. 
  Therefore, a high misdecoding probability at the first index of a repetition 
  block has a more fatal influence than an unreliable decision at the end. 
  Consequently, a repetition block should always start with the most reliable bit channel. 
  Blocks of larger length are built in an analog fashion. Here, also the most reliable bit channel should be put in front. 

  Finally, the rate of the original code has to be preserved, which leads to further obvious restrictions on the number of possible repetition blocks. 
  Finding the optimum from the remaining possible outer coding schemes is easily accomplished by an exhaustive search. 
  
  Fig.~\ref{fig_bitch} shows an example of an outer coding scheme constructed according to the above-mentioned constraints. 
  The markers represent the bit channels used by the concatenated code. Here, the repetition blocks are visualized by red lines, 
  the red markers denote the corresponding equivalent bit channels while the black markers stand for the unmodified bit channels of the concatenated code. 
  We remark that further increasing the rate $R_\mathrm{i}$ of the inner code has no significant effect on the performance.  

  Clearly, the proposed scheme can easily be extended to using higher-dimensional outer codes for increased performance, 
  as considered in \cite{Seidl:10}, though at the cost of an increased complexity. Moreover, the results from~\cite{Seidl:10} indicate that the possible 
  additional gain will not be large.

\subsection{Decoding}
  \begin{figure}[width=0.48\textwidth]
    \begin{centering} %
  \beginpgfgraphicnamed{pic_rep-external}%
  \input{pic_rep.tikz}%
  \endpgfgraphicnamed%
 \end{centering}
    \caption{\label{algo_inner}SC decoding of an outer repetition code operating on $u_0$ and $u_3$. Bold face lines: Inspected (and tentatively selected) paths in the SC decoding process. Dashed lines: 
              Inspected but discarded paths. Thin lines: Never inspected paths.}
  \end{figure}
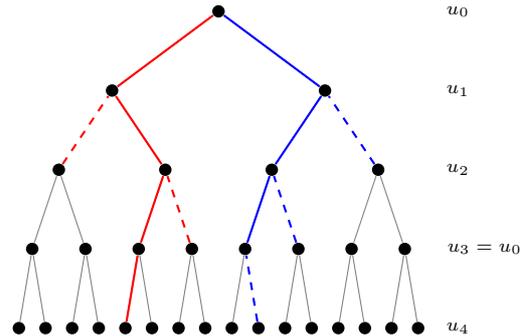
  For joint decoding of inner and outer code, we apply the original SC algorithm with a slight modification, only. 
  Decoding of a received vector starts as usual. 
  Assume now that two source bits $u_i$ and $u_j$ are protected by a repetition code, i.e., $u_i=u_j$ for some $j>i$. 
  On reaching stage $i$, instead of taking a hard decision on $u_i$, the decoder creates a new branch and tests 
  both possibilities by determining the sequences 
  \begin{align*}
    \ve{s}_0 &= \langle 0,\hat u_{0,i+1}\cdots \hat u_{0,j-1},0\rangle \; , \\
    \ve{s}_1 &= \langle 1,\hat u_{1,i+1}\cdots \hat u_{1,j-1},1\rangle \; .
  \end{align*}
  For decisions on the symbols $\hat u_{0, i+1}, \ldots , \hat u_{0, j-1}$ and $\hat u_{1, i+1}, \ldots , \hat u_{1, j-1}$, the conventional SC decision rule~\eqref{scdec} is applied. 
  Afterwards, the more likely of the two sequences is selected:
  \begin{equation}
     \langle \hat u_i\cdots \hat u_j\rangle := \ve{s}_{b^{\ast}}
  \end{equation}
  where
  \begin{equation}\label{innerdec}
    b^{\ast} = \argmax_{b \in \{0,1\}} \Big\{ \mathrm{Pr}\big(\ve{S}_b = \ve{s}_b | \ve{Y}, \hat U_0 \cdots \hat U_{i-1},\ve{S}_b\big) \Big\}\;. 
  \end{equation}
  The other path is discarded. The decoding scheme is visualized in Fig.~\ref{algo_inner} for a simple example code with $u_0=u_3$. 
  Repetition codes of larger length are decoded in an analog fashion. 
  Clearly, the decoding complexity is at most doubled since we exclude overlapping blocks. Furthermore, the decoding of outer repetition codes in this way can easily be integrated 
  into improved versions of the SC decoder, e.g., list decoding. 

%% file: pic_rep.tikz
\begin{tikzpicture}
[lineDecorate/.style={-, color=gray},%
lineDecorate2/.style={-,thick,color=red},%
lineDecorate4/.style={-,dashed, thick,color=red},%
lineDecorate3/.style={-,thick,color=blue},%
lineDecorate5/.style={-,dashed, thick,color=blue},%
VarNode/.style={shape=circle,fill=black, thin, inner sep=0pt},%
CheckNode/.style={shape=rectangle,inner sep=2.5pt,draw,thick}, scale=0.35]
\scriptsize
\foreach \nodename/\nodelabel/\x/\y/\direction/\navigate in {
u40/c/0/0/bottom/south, u41/c/1/0/bottom/south, u42/c/2/0/bottom/south, u43/c/3/0/bottom/south, 
u44/c/4/0/bottom/south, u45/c/5/0/bottom/south, u46/c/6/0/bottom/south, u47/c/7/0/bottom/south, 
u48/c/8/0/bottom/south, u49/c/9/0/bottom/south, u410/c/10/0/bottom/south, u411/c/11/0/bottom/south, 
u412/c/12/0/bottom/south, u413/c/13/0/bottom/south, u414/c/14/0/bottom/south, u415/c/15/0/bottom/south, 
u30/c/0.5/3/bottom/south, u31/c/2.5/3/bottom/south, u32/c/4.5/3/bottom/south, u33/c/6.5/3/bottom/south, 
u34/c/8.5/3/bottom/south, u35/c/10.5/3/bottom/south, u36/c/12.5/3/bottom/south, u37/c/14.5/3/bottom/south, 
u20/c/1.5/6/bottom/south, u21/c/5.5/6/bottom/south, u22/c/9.5/6/bottom/south, u23/c/13.5/6/bottom/south, 
u10/c/3.5/9/bottom/south, u11/c/11.5/9/bottom/south, 
u00/c/7.5/12/bottom/south}
{ \node (\nodename) at (\x,\y) [VarNode] {$\nodelabel$}; }
\path
\foreach \startnode/\endnode in {
u40/u30, u41/u30, u42/u31, u43/u31, u44/u32, u45/u32, u46/u33, u47/u33, u48/u34, 
u410/u35, u411/u35, u412/u36, u413/u36, u414/u37, u415/u37, 
u30/u20, u31/u20, u32/u21, u34/u22, u36/u23, u37/u23, 
u21/u10, u22/u11, u10/u00, u11/u00}
{ (\startnode) edge[lineDecorate] (\endnode) };


\path
\foreach \startnode/\endnode in {u32/u21, u21/u10, u44/u32, u10/u00}
{ (\startnode) edge[lineDecorate2] (\endnode) };
\path
\foreach \startnode/\endnode in {u11/u00, u34/u22, u22/u11}
{ (\startnode) edge[lineDecorate3] (\endnode) };
\path
\foreach \startnode/\endnode in {u33/u21, u20/u10}
{ (\startnode) edge[lineDecorate4] (\endnode) };
\path
\foreach \startnode/\endnode in {u49/u34, u23/u11, u35/u22}
{ (\startnode) edge[lineDecorate5] (\endnode) };
\node (u0) at (16.5,12) {$u_0$};
\node (u1) at (16.5, 9) {$u_1$};
\node (u2) at (16.5, 6) {$u_2$};
\node (u3) at (17.5, 3) {$u_3=u_0$};
\node (u4) at (16.5, 0) {$u_4$};
\end{tikzpicture}

%% file: sec_sim.tex
\begin{figure}
  \psfrag{ebno}[Bc][Bc][0.8]{$10\log_{10}(E_\mathrm{b}/N_\mathrm{0})$}
  \psfrag{wer}[Bc][Bc][0.8]{$\mathrm{WER}$}
  \centerline{\includegraphics[width=0.48\textwidth]{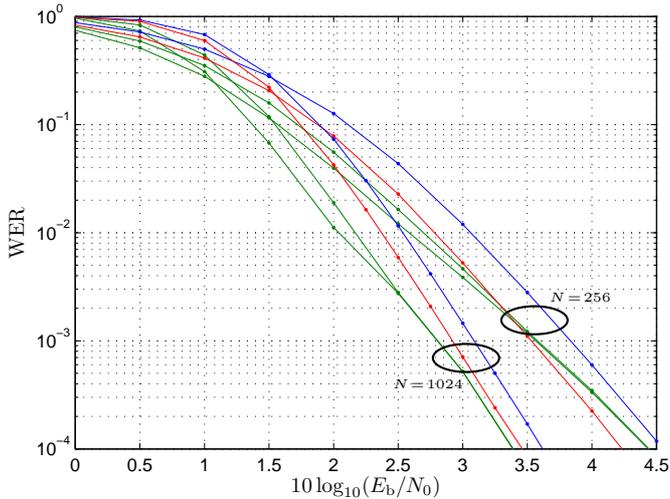}\psellipse(-1.85,2.4)(.45,.2)\psellipse(-2.75,1.9)(.45,.2)%
  \rput(-1.25,2.7){\tiny $N\!=\!256$}\rput(-3.25,1.55){\tiny $N\!=\!1024$}}
  \caption{\label{fig_sim}Simulation results: BPSK-AWGN channel, polar code block length $N=256, 1024$, rate $R=1/2$. Blue: SC decoding. Green: successive list decoding (list size $L=2,4$). Red: proposed concatenation scheme.}
\end{figure}
Fig.~\ref{fig_sim} shows simulation results for polar coding schemes of block length $N=256$ and $N=1024$ transmitted over a BPSK-AWGN channel. 
The shorter and longer code have been optimized (according to \eqref{sc_wer}) for $\Eb/\No=2.5\;\dB$ and $\Eb/\No=2.0\;\dB$, respectively. 

Compared to the original, SC-decoded polar codes (blue), the use of an improved decoder like the successive list decoder (green) shows an SNR-dependent effect: 
In the low-SNR region, significant gains are achieved while for increasing SNR the performance advantage vanishes. Here, both decoders perform close to ML decoding. 

The proposed scheme (red) leads to an improved performance in a similarly efficient way with a complexity comparable to that of a list decoder with $L=2$. 
However, in contrast to an optimized decoder, it achieves a constant coding gain 
of approx. $0.3\;\dB$ ($0.2\;\dB$ for the longer code) over the SC-decoded code at all SNR regimes. 
Therefore, at high SNR it is able to outperform even a list decoder with large list size or an ML decoder, but with much lower complexity, because the rate- and length-preserving 
concatenation yields an improved code. 

%% file: sec_conc.tex
The proposed concatenation scheme may be seen as a method to overcome the quantization effect when constructing a polar code that is caused 
by a hard selection of the bit channels (each channel is either used for information transmission or frozen). \newpage
As this quantization vanishes with increasing block length and polarization, the scheme is certainly restricted to polar codes of short to moderate length. 
Although the achievable performance gain is not too large, it comes at  
very small additional costs. 
When used together with an improved polar decoder, the 
beneficial effects of both approaches are combined. 
Furthermore, the proposed scheme can itself be used as an inner code in 
other concatenation approaches -- at least if inner and outer decoding are performed separately there like in~\cite{Seidl:11}. 
In this case, the coding gain in error performance is preserved. 